\documentclass[12pt]{iopart}
\usepackage{graphicx}

\begin{document}

\title[ ]{Evidence for the predominant influence of the asymmetry degree of
freedom on the even-odd structure in fission-fragment yields
}

\author{Manuel Caama\~{n}o $^1$ \footnote{Present address: Universidade de Santiago de Compostela, E-15706 Santiago de Compostela, Spain}~, Fanny Rejmund$^1$ and Karl-Heinz~Schmidt $^2$}

\address{$^1$GANIL, CNRS/IN2P3, CEA/DSM, bd Henri Becquerel, 14076 Caen}

\address{$^2$GSI, Planckstrasse 1, 64291 Darmstadt, Germany}

\ead{frejmund@ganil.fr}

\begin{abstract}
Based on a wide systematics of fission-fragment distributions measured in low-energy fission, the even-odd staggering in the fission-fragment element yields is investigated. The well-established evolution of the global even-odd effect with the fissioning system is found to be only a partial aspect of the even-odd structure. Indeed, it is shown that the global even-odd effect is varying systematically with the mean asymmetry of the fission-fragment distribution, and that the general increase of the even-odd staggering with  asymmetry is depending on the fissioning system. Thus, 
the dependency of the even-odd effect with the fissioning system is accredited in part to the asymmetry evolution of the charge distribution, and not solely related to the dissipated energy as it has been done earlier. This interpretation is strongly supported by data measured in inverse kinematics, which cover the complete charge distribution and include precise yields at symmetry. The relevance of the order parameter to describe the even-odd effect in fission-fragment yields as a general property is explored.

\end{abstract}

\pacs{ 24.75.+i, 25.70.-z, 25.85.-w, 25.85.Ec, 25.85.Jg}


\section{Introduction}

Even-odd staggering in proton (Z) and neutron (N) numbers is a well-known characteristic of low-energy fission-fragment yields \cite{amiel}. 
The large amplitude of this staggering, which may reach 40\% in the case of the element distribution of fission fragments of thorium \cite{bocquet}, has always been fascinating to nuclear physicists. 
The experimental access to the exact number of neutrons of the fission fragments at scission is difficult to obtain,  since the number of neutrons is modified
by evaporation processes. In contrast, evaporation of protons from the neutron-rich fission fragments is negligible. Therefore, if measured prior their beta-decay, the  element yields, and the associated even-odd staggering, reveal the properties of the fission split at scission.
The observation of odd-Z fragments from an even-Z fissioning nucleus testifies the breaking of proton pairs during the re-organisation of the intrinsic structure of the fissioning nucleus on its path towards scission. Indeed, in thermal-neutron-induced fission of even-Z actinides, the compound nucleus reaches the saddle deformation with an intrinsic excitation energy below the pairing gap. Thus, an ensemble of fully paired protons undergoes the deformation down to scission, where at least one pair is broken and both unpaired protons end up in different fission fragments to produce odd-Z species. 
The amplitude of the even-odd staggering is therefore linked to dissipation during the deformation in the fission process. It has always been a challenge to understand the mechanism of dissipation and the relation between pairing effects and the dissipated energy. The present work focuses on data for which nuclear-charge yields are accessible. They emerge from experiments performed at ILL for fissile actinides and at GSI for long chains of neutron-deficient thorium and uranium isotopes. This ensemble of data allows accessing an unprecedented systematic on pairing effects in fission-fragment yields. They show new general properties that cannot be explained with the conventional interpretations.
 
\section{Global even-odd staggering and dissipated energy}
\label{globalandissipated}
A simple method to quantify the even-odd structure in the fragment production is to define the global even-odd effect $\delta$  as the average difference between even- and odd-Z yields over the full available distribution \cite{bocquet}.
Several models have attempted to quantitatively relate the global even-odd staggering to the dissipated energy. They are all based on the consideration that the even-odd effect in element yields is determined by the probability that the protons remain in a fully paired configuration at scission. 

The model proposed by Nifenecker et al. \cite{nif} is certainly the most widely used as it correlates   $\delta$ and the dissipated energy $E_{diss}$ with a very simple expression. This model considers a maximum number of broken pairs at scission, which is determined as the ratio between the dissipated energy and the amount of energy necessary to break a pair (the pairing gap). The nucleus at scission is then considered as a fully paired core and an ensemble of broken pairs, to which a combinatorial analysis is applied in order to determine the probability to break a proton versus a neutron pair, the probability to break a pair if the necessary amount of energy is gained, and finally the probability that the nucleons of the broken pair end up in two different fragments. This model estimates a dissipated energy of about 4 MeV for $^{230}$Th to 12 MeV for $^{250}$Cf. 

An alternative approach by Bouzid et al. \cite{bouzid} is based on dynamical considerations of the fission process. In this model, the descent from saddle to scission is considered to be adiabatic, and the violent neck rupture leading to the formation of the two separated fragments causes the pair breaking. The probability to break a pair is correlated with the velocity of the neck rupture, which is shown to increase with the Coulomb repulsion at the scission point. The Coulomb repulsion is linked to the Coulomb parameter  $Z_c = {Z_{f}^2}/{A_{f}^{1/3}}$, where $Z_f$ and $A_f$ are the nuclear charge and mass of the fissioning nucleus. The probability that the broken pairs end up in different fragments is a parameter fitted to reproduce the data. In particular, this model considers the difference in the proton- and neutron-number staggering as a consequence of the less violent neck rupture for protons, since they are less present in the neck due to the Coulomb repulsion.

A rigorous formulation  of the even-odd staggering has been derived by Rejmund et al. \cite{rejmund} in the frame of the statistical model. It is based on a realistic description of the number of quasi-particle excitations of the proton and neutron sub-systems as a function of the excitation energy at scission. The pairing-gap parameter depends on excitation energy, deformation,  and the number of quasi-particle excitations. The probability to break proton pairs as a function of the excitation energy is derived accurately as the ratio of the proton excited states over all available single-particle states. This is in contrast to the model of Nifenecker et al., where the even-odd effect is deduced from a combinatorial analysis, based on the ensemble of possibly broken pairs, using fitted parameters. 

\begin{figure}
\hspace{-2cm}
\includegraphics[height=.30\textheight]{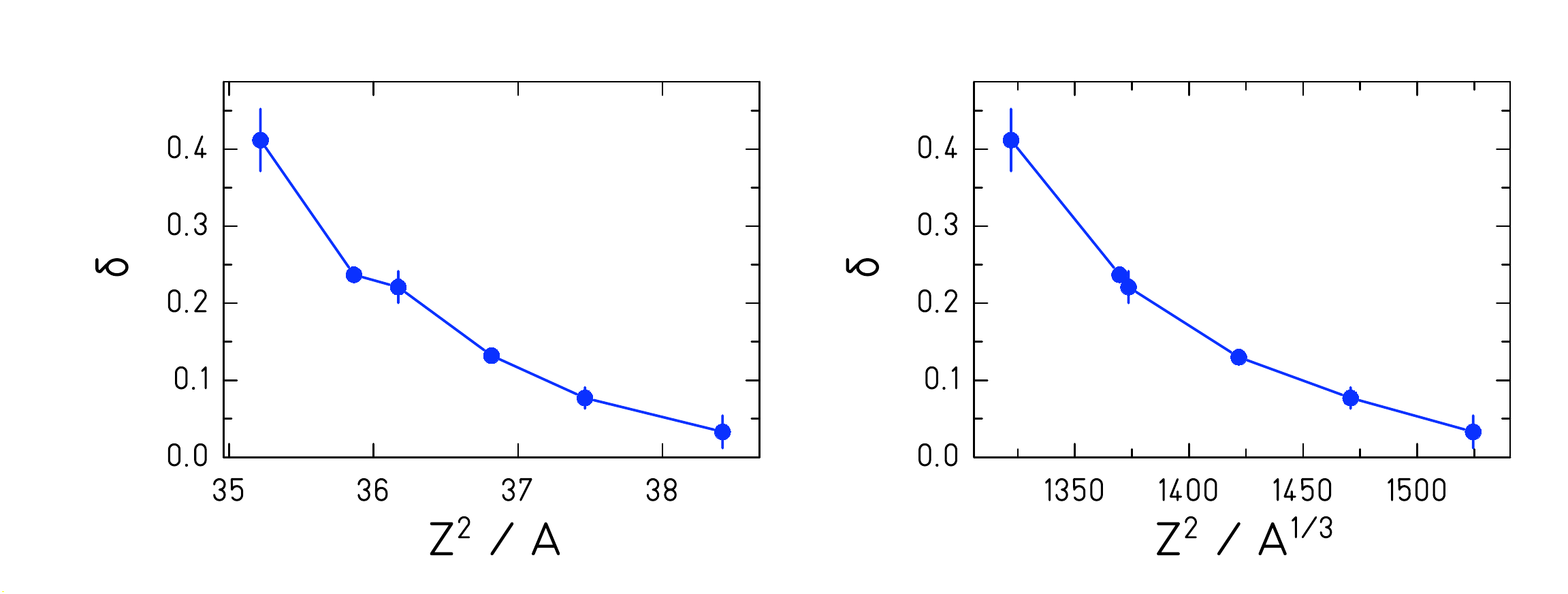}
 \caption{Global even-odd effect as a function of the fissility parameter (left) or the Coulomb parameter (right). 
  Data are from thermal-neutron-induced fission of $^{229}$Th \cite{bocquet}, $^{233}$U \cite{quade}, $^{235}$U \cite{lang, sida}, $^{239}$Pu \cite{schmitt},   $^{245}$Cm \cite{rochman02, rochman04} and $^{249}$Cf \cite{djebara89,hentzschel}. }
  \label{f1}
\end{figure}

In the measurements of fragment yields of fissile actinides at ILL, it has been observed that the global even-odd effect decreases with the fissility of the fissioning system \cite{goennenwein-wage}. The amplitude of the even-odd effect being associated to the dissipated energy gained by the nucleus, the decrease of the even-odd staggering with fissility has brought up the idea that more energy is dissipated in the descent from saddle to scission as the fissility of the fissioning nucleus increases. The fissility parameter $x = {Z_{f}^2}/{A_{f}}$ reflects the stability against fission, but its connection with the dissipated energy is not clear. 
On the other hand, the total energy release from saddle to scission may be expressed with a linear correlation with the Coulomb parameter $Z_c = {Z_{f}^2}/{A_{f}^{1/3}}$  \cite{bouzid} and therefore, the evolution of the even-odd staggering with the fissioning nucleus is usually investigated as a function of the Coulomb parameter \cite{ rejmund, bocquet2, naik}. 
The evolution of the global even-odd staggering is displayed as a function of the fissility and the Coulomb parameter in Fig. \ref{f1}, left and right panels, respectively. A clear decrease is observed as a function of both parameters.

\section{Evolution of the  fragment distribution with the fissioning nucleus and its influence on the even-odd staggering}

\begin{figure}
 \includegraphics[height=.7\textheight]{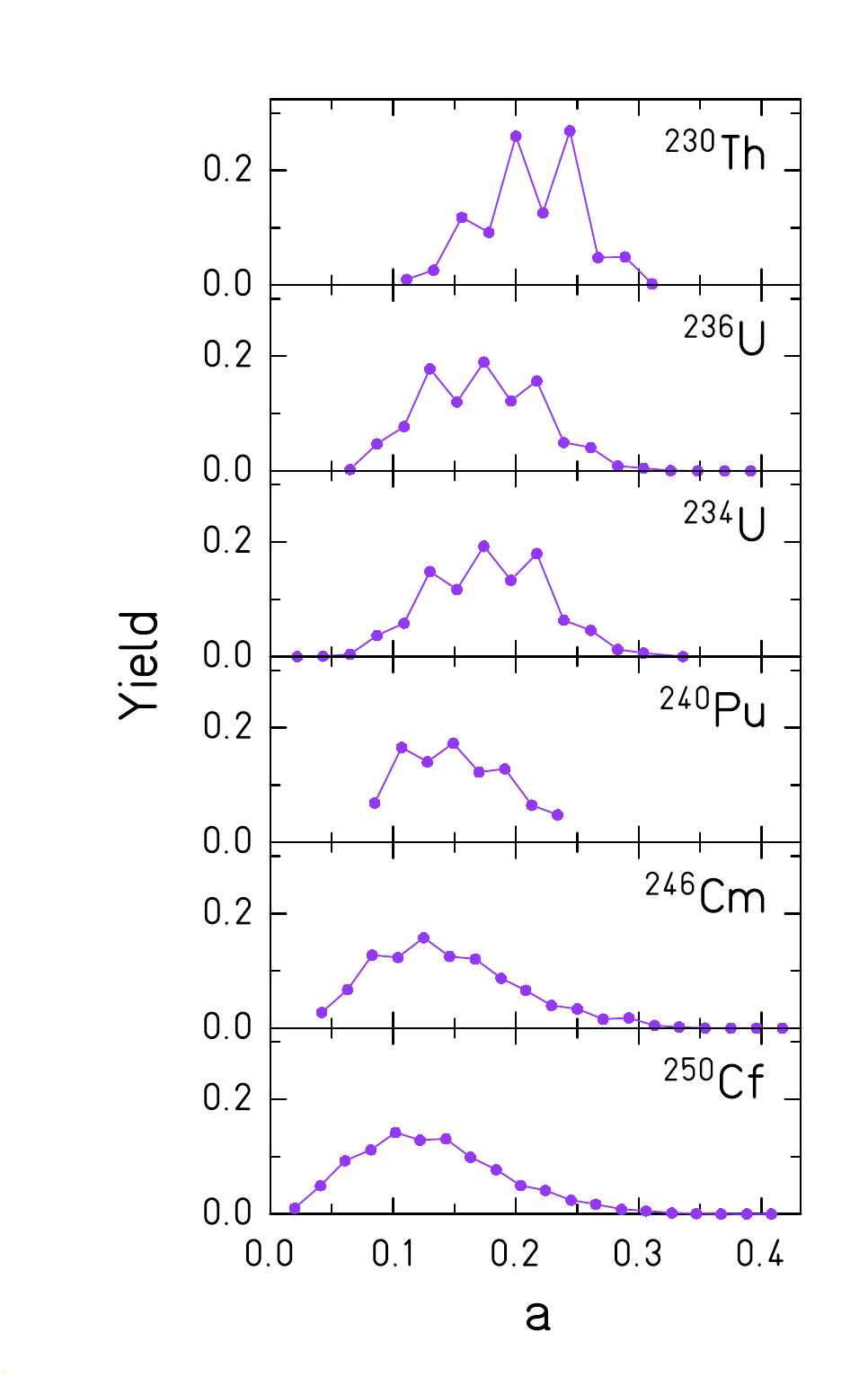}
  \caption{Element distributions displayed as a function of the asymmetry parameter of the distribution, for thermal-neutron-induced fission of $^{230}$Th, $^{236}$U, $^{234}$U, $^{240}$Pu, $^{246}$Cm and $^{250}$Cf, from top to bottom. }
 \label{distz}
\end{figure}

\begin{figure}
 \includegraphics[height=.3\textheight]{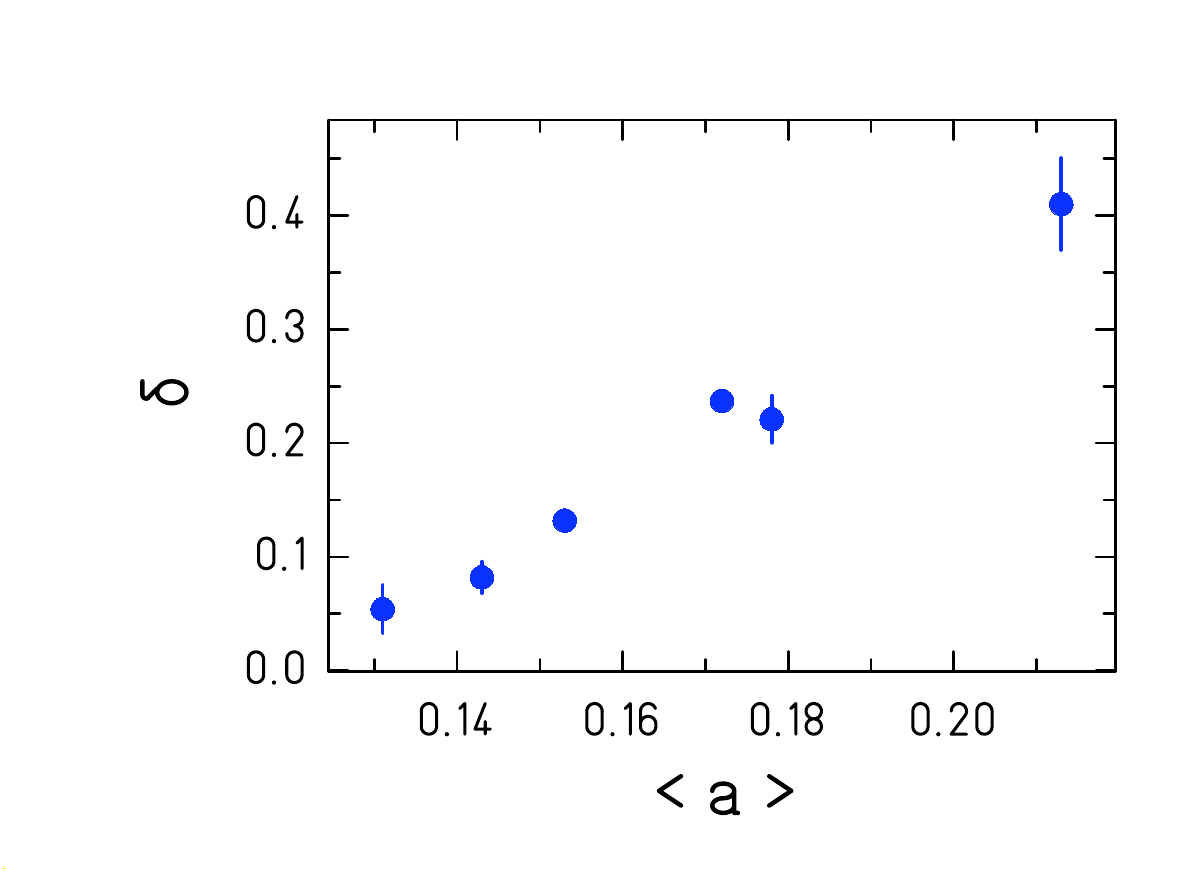}
  \caption{Global even-odd effect as a function of the average asymmetry of the fragment distribution.}
 \label{delta_a}
\end{figure}

In low-energy fission of most actinides, shell effects induce an asymmetry in the mass and charge distributions of fragments, which show two groups \cite{unik}. The group of heavy fragments is distributed over an average value of A$\sim$140, independently of the mass of the fissioning system.  In order to keep the total mass conservation, the group of the light fragments moves towards heavier masses for heavier fissioning nucleus, approaching gradually the symmetry.  The equivalent behaviour is observed in charge distributions. However, in this last case, until the recent application of the inverse kinematics technique \cite{schmidt}, only the light part of the distribution was accessible experimentally.
Element-yield distributions reported in figure \ref{distz} have been obtained  
at the Lohengrin spectrometer  \cite{bocquet,  quade,  lang, sida, schmitt, rochman02, rochman04, djebara89, hentzschel}. For this particular experimental procedure, the element yields for which the complete isotopic distribution could be measured have been selected. This restriction is necessary to avoid incorrect element yields  \cite{rejmund2}. They show an average light charge centred on values varying from Z = 36 to 44, when considering fissioning nuclei from Th to Cf, as pointed out with an arrow in figure \ref{combi}. This coincides with a heavy charge distribution centred on a constant value Z=54, independently of the fissioning nucleus, as was already reported in \cite{boeck}. To explore the fission asymmetry, it is convenient to define the asymmetry parameter:
$
a = {(Z_H-Z_L)}/{Z_f}
$
, where $Z_H$ and $Z_L$ are the charge of heavy and light fragment, respectively. In figure \ref{distz}, the element yields are displayed as a function of this parameter, which results in a common scale on the asymmetry of the distribution for all systems considered. The distribution of the light fragments is gradually approaching the symmetry, as the fissioning system is getting heavier. Effectively, for the fissile systems investigated, the average asymmetry $ <a>$ of the nuclear-charge distributions of the light fission-fragments shifts from 0.28 to 0.13 when considering fissioning systems from $^{230}$Th to $^{250}$Cf. In figure \ref{delta_a},  the corresponding global even-odd effect is displayed as a function of the average asymmetry of the light fragment distribution. A clear relationship is observed; when the average asymmetry decreases, the global even-odd effect decreases. 

As there exists a clear decrease of the even-odd effect  with increasing excitation energy at saddle \cite{pommŽ, persyn} and a prediction for an increasing energy release between saddle and scission with the Coulomb parameter \cite{bouzid, bocquet2}, it is conventionally assumed that the decrease in even-odd effect with the fissility or with the Coulomb parameter is due to the increase in dissipated energy from saddle to scission. However, figures \ref{f1} and \ref{delta_a} show that the global even-odd effect is correlated also with other parameters which are not manifestly connected to the energy dissipated between saddle and scission; the fissility $x$ and the average asymmetry $< a >$of the fragment distribution. The clear correlation between the global even-odd effect and the mean asymmetry opens the question on a possible origin in the asymmetry degree of freedom, of a substantial contribution to the even-odd effect. 
In the following subsections, the even-odd staggering of
the element distribution as a function of the charge split is investigated,  in order to get more detailed information on the influence of the asymmetry degree of
freedom on the magnitude of the even-odd effect.

\subsection{Local even-odd staggering and influence of the asymmetry}

\begin{figure} 
\includegraphics[height=.45\textheight]{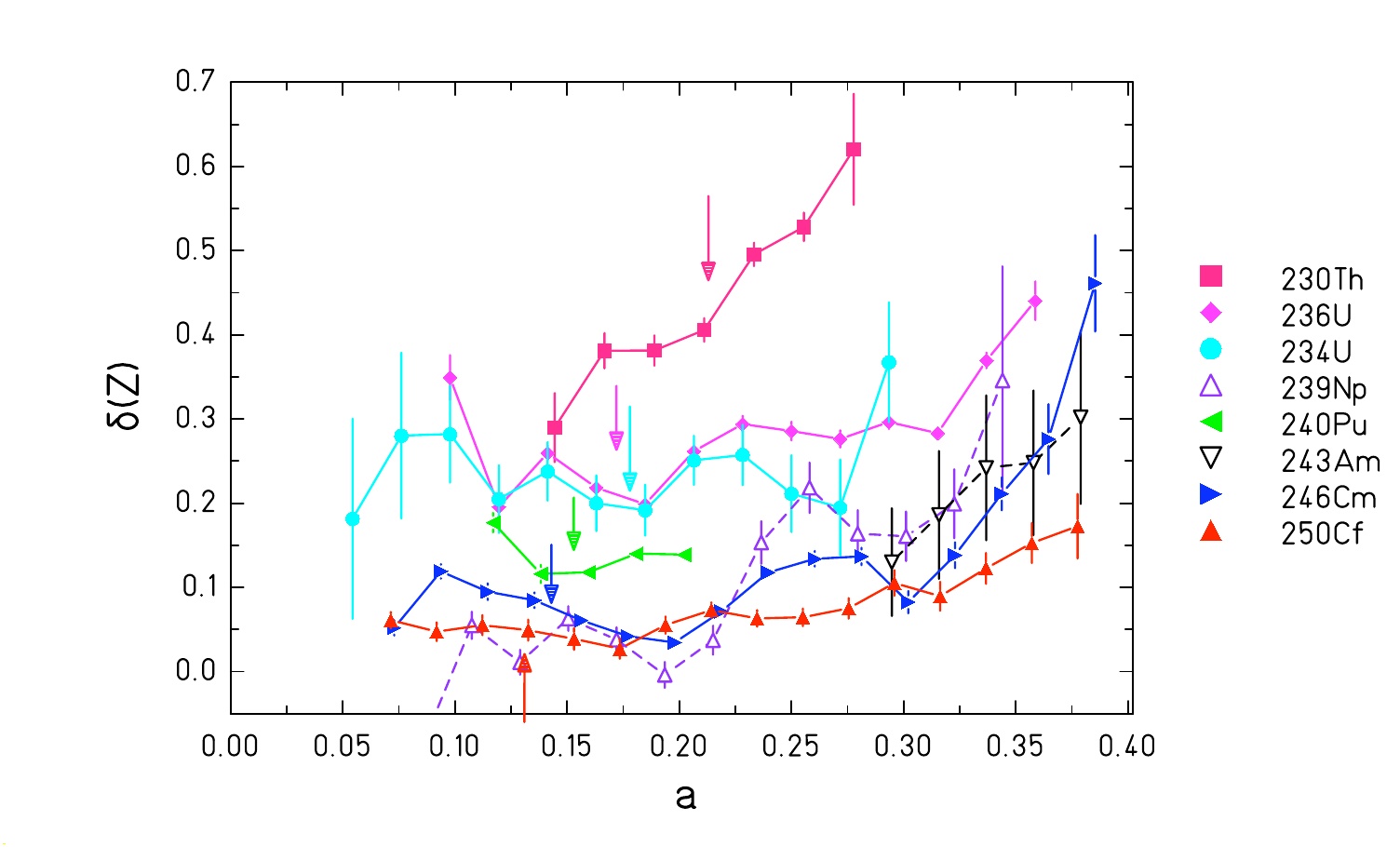}
\caption{Local even-odd effect for the series of element distributions of even- \cite{ bocquet, quade,lang,  sida,  schmitt, rochman02, rochman04,  djebara89, hentzschel} and odd-Z \cite{tsekhanovich01, tsekhanovich99} fissile nuclei. The arrows indicate the value of the average asymmetry of the fragment distributions. }
\label{combi}
\end{figure}

The local even-odd staggering $\delta(Z)$ is a measure of the deviation of the nuclear-charge distribution from a smooth behaviour, and is usually studied following the prescription of Tracy et al. \cite{tracy}. For different fissioning systems \cite{bocquet, quade, lang, sida, schmitt, rochman02, rochman04,  djebara89, hentzschel}, the even-odd staggering has been  shown to be larger for large asymmetry. This experimental observation is reproduced in figure \ref{combi} in a comprehensive view.
The rate of increase is largest for the lightest system $^{230}$Th and decreases for the heavier systems. 
The values of the even-odd effect for the different systems approach each other at symmetry, but due to missing
experimental information at symmetry and due to the scattering of the data points one can only say
that they converge to a rather narrow range between 5 and 20 percent. 
For several systems, the data points  that are closest to symmetry present appreciably higher local even-odd effect than
expected from the global trend. This effect may be associated to the influence of the Z=50 shell in
the complementary fragment, which is known to enhance the yield of tin isotopes and, thus, leads to a local increase of the deduced even-odd effect, because its derivation cannot distinguish between structures caused by pairing or shell effects.
The general trend of increase towards asymmetry has led to the notion of cold asymmetric fission, for which extreme deformation would take most of the available excitation energy, and consequently the intrinsic excitation energy would remain low \cite{sida}. However, no elaborate model exists to quantitatively describe the even-odd effect based on these assumptions. Moreover, it will be shown in the following that even in the case of scission configurations with invariably at least one unpaired proton, large even-odd effect can be observed at large asymmetry. Therefore, the excitation energy and the consequent number of broken pairs cannot be considered as uniquely accountable for the amplitude of the even-odd effect.



\subsection{Local even-odd staggering in odd-Z fissioning systems}

In the late 90\'{}s the general appearance of a large even-odd effect in the Z distributions of odd-Z fissioning nuclei  has been reported \cite{stein}. In these odd-Z systems, the probability to have at least one unpaired proton is always one. Assuming that any unpaired proton ends up in one or the other fragment with equal probability, it was expected that the even-odd effect would be zero over the full element distribution. With experimental techniques based on inverse kinematics, the even-odd staggering has been measured for a large systematics of actinium and protactinium isotopes, over the complete fragment distribution. Its value was found to be zero close to symmetry and systematically increasing for large asymmetry, up to amplitudes as large as 40 \%. 
A similar observation of large even-odd effect has been reported in neutron-induced fission of $^{238}$Np \cite{tsekhanovich01},  and $^{242m}$Am \cite{tsekhanovich99}, as shown in figure \ref{combi}. The even-odd effect of the odd-Z fissioning systems at large asymmetry
follows the general behaviour of the even-Z fissioning systems. Their values are very close to those
of the neighbouring even-Z systems, as was reported already in \cite{tsekhanovich01}. 
In the inverse kinematics data, where the heavier part of the fragment distribution has been measured, the even-odd staggering was found to be negative, revealing a higher probability for the unpaired proton to end up in the heavy fragment. 
Since there is no preference for unpaired protons to end up in one or the other fragment in symmetric fission by trivial reasons, the even-odd
effect in odd-Z fissioning systems is necessarily zero for symmetric
splits. In these odd-Z fissioning systems, an even-odd effect may only be generated by a mechanism which is related to some
difference in the properties of the fragments in asymmetric splits. The similarity of the values
of close-by even-Z and odd-Z systems suggests that a similar mechanism is responsible for the increase of the even-odd effect in even-Z fissioning systems.
These experimental observations collected in this work lead to the interpretation that the conventionally assumed  relation between pair breaking and  even-odd effect in the fission yields is valid only close to symmetry, and that it is limited to a restricted value below 20\%, which does not vary much as the fissility (or Coulomb parameter) of the fissioning system varies. Most of the amplitude of the even-odd effect is then linked to the asymmetry component of the fragment distribution, and is common to the odd-Z and even-Z fissioning systems. 

The observation of an even-odd staggering for odd-Z fissioning nuclei and its interpretation reveal that the relation between the amplitude of the even-odd staggering in fission-fragment charge yields and the intrinsic excitation energy at scission is not as direct as suggested by the models discussed in section \ref{globalandissipated}. Indeed, neither the statistical description of Nifenecker et al. nor the dynamical description of Bouzid et al. can explain the appearance of an even-odd structure for odd-Z fissioning nuclei since in these models, the probability of unpaired nucleons to end up in one or the other fragment does not depend on the size of the fragments. The statistical model of Rejmund et al. \cite{rejmund} considers the effect of dissipated energy at  symmetric scission,  separately of the influence of asymmetry in the fission-fragment yields. 
In references \cite{tsekhanovich01, tsekhanovich99}, the phenomenon is qualitatively discussed in terms of energy balance; as the pairing gap is decreasing with the fragment mass, the light fragments remain paired preferably. However, only the binding energy of a cold system is considered, without any intrinsic excitation. A statistical description of the even-odd staggering with the asymmetry based on the level density of the fission fragments formed at scission  \cite{stein} reproduces the larger probability for the unpaired nucleons to end up in the heavy fragment. This model gives a quantitative prediction of the general increase of the even-odd staggering with the asymmetry, for odd-Z fissioning nuclei as well as for even-Z fissioning nuclei.  
Recently, a mechanism has been proposed that induces the large even-odd effect in asymmetric splits by the energy-sorting process in superfluid nuclear dynamics \cite{jurado}.

\subsection{Local even-odd staggering at symmetry}

\begin{figure}[htbp] 

    \hspace{-2.5cm}\includegraphics[height=0.3\textheight]{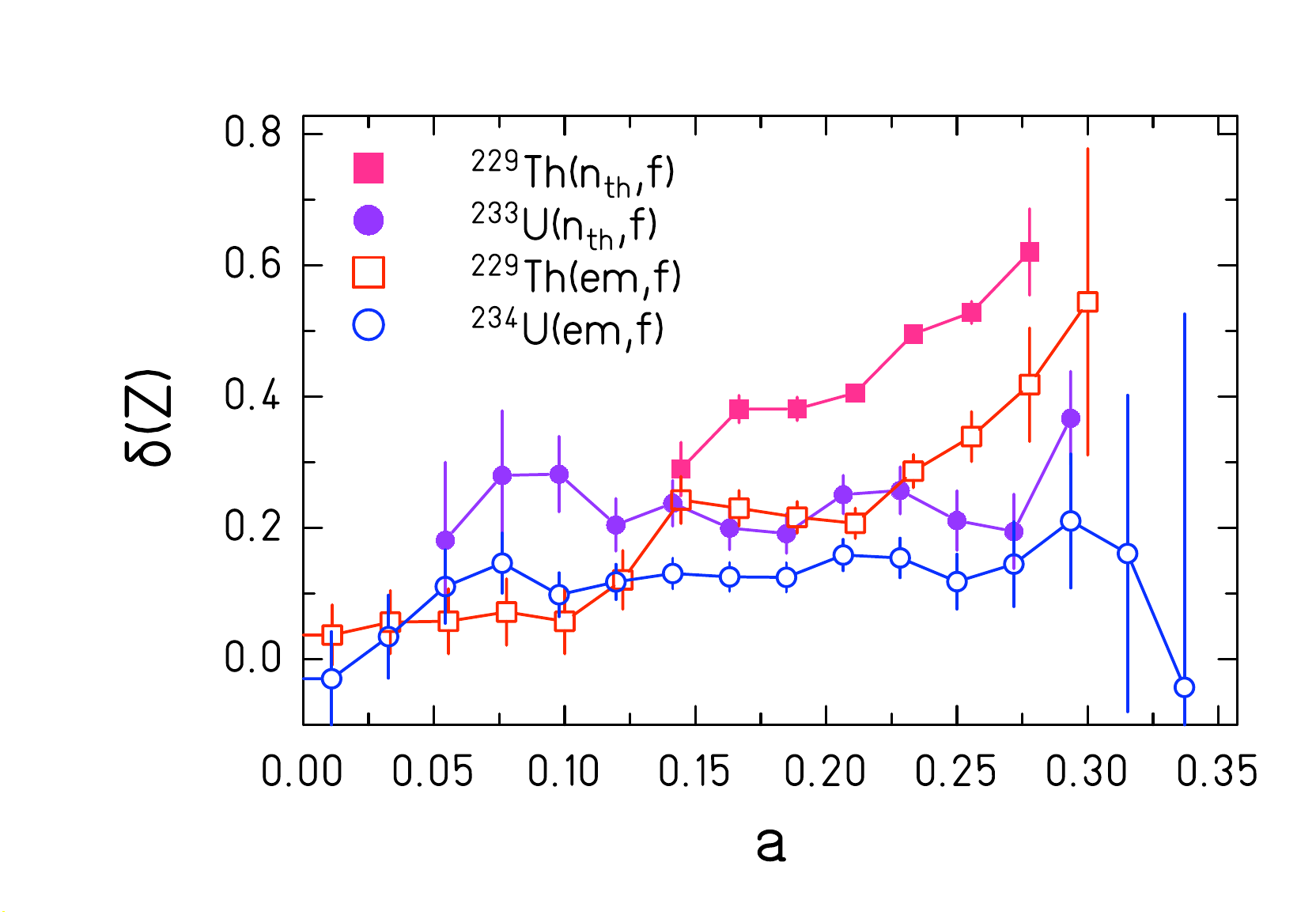} \hspace{-1.cm}\includegraphics[height=0.3\textheight]{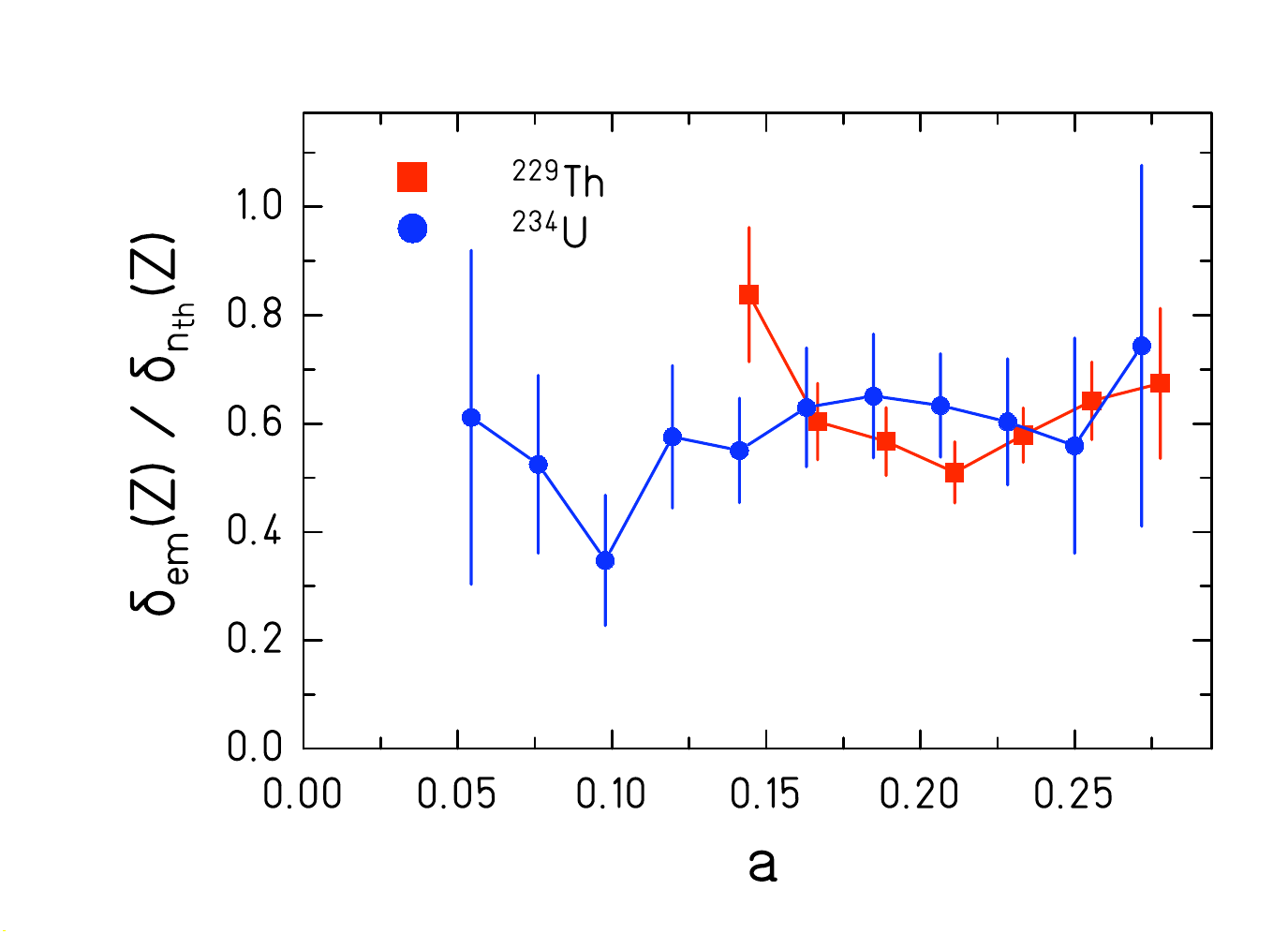} 
   \caption{Left panel: Local even-odd staggering measured in similar fissioning systems using thermal-neutron- (full symbols) and electromagnetic-induced- (open symbols) fission. Right panel: Ratio between the two measures for the two fissioning systems considered.}
   \label{delta_sym}
\end{figure}

The general trend of the even-odd effect towards symmetry is to decrease and to remain restricted to the range between 5 and 20\%, as depicted in figure \ref{combi}. However, due to the lack of data close to symmetry, it is difficult to conclude further. The experimental technique based on inverse kinematics \cite{schmidt} allows to access the element yields over the complete element distribution, with high resolution and for a wide systematic of neutron-deficient fissioning systems. In addition, fission is induced in electromagnetic interaction, which excites few MeV above the thermal-neutron capture, resulting in higher yields for symmetric splits. As a consequence, this technique provides unique and high-quality data on element yields at symmetry.
In figure \ref{delta_sym}, the local even-odd effect measured in similar fissioning systems using both techniques (thermal-neutron-induced fission in direct kinematics and electromagnetic induced fission in inverse kinematics) is displayed in left panel. The inverse-kinematics data cover the full element distribution; they present a general lower value, resulting from the higher excitation energy of the compound nucleus, but preserve similar characteristic patterns as observed in thermal-neutron-induced fission. This is demonstrated in the right panel of figure \ref{delta_sym}, where the ratio between the amplitude of the even-odd staggering in both experiments is displayed and  shows a constant value around 60\%. From this ratio, it is possible to extrapolate that for both fissioning systems ($^{230}$Th and $^{234}$U), the even-odd effect in low energy fission at symmetry is below 10\%, of the order of 7\%. It is important to note, that this small but non-zero value of the even-odd staggering at symmetry has two major consequences. Firstly, the parameterization used for the evaluation of fission-fragment yields \cite{wahl}, which is based on former radio-chemical data, considers no even-odd effect at symmetry, in contradiction to the present experimental observation. The parameterization should be reformulated appropriately.

Secondly, the major contribution to the global even-odd effect (40\% for $^{230}$Th, 23 \% for $^{234}$U, see figure \ref{f1}) emanates from the asymmetry of the fragment distribution, which is not directly related to the probability to  break pairs in the descent from saddle to scission, as it is also present in odd-Z fissioning systems. As a consequence, the previous estimations on the dissipated energy derived from the amplitude of the global even-odd staggering \cite{nif, bouzid} are not justified.
In addition, the estimated value for the even-odd effect at symmetry for the two systems $^{230}$Th and $^{234}$U is very close to the value measured close to symmetry for heavier systems  as $^{246}$Cm and $^{250}$Cf, which are also of the order of few \% (see figure \ref{combi}). Within the experimental uncertainty, the variation of the amplitude of the even-odd effect at symmetry with the fissility of the fissioning nucleus shows to be weak. The standard interpretation of the even-odd effect attributes a strong variation of the dissipated energy  with the fissility of the fissioning system, based on the evolution of the global even-odd effect with the fissility (figure \ref{f1}). The present investigation shows that, in reality it is the asymmetry-based component of the even-odd effect which undergoes an important variation with the fissility of the fissioning system. The even-odd staggering at symmetry remains in a range of small values, and shows a weak if no variation within the experimental error bars. However, due to the lack of data at symmetry, no definite conclusion on the variation of the even-odd effect at symmetry in fissile nuclei can be drawn.

\begin{figure} 
\hspace{-1cm}
 \hspace{-1cm}\includegraphics[height=.7\textheight]{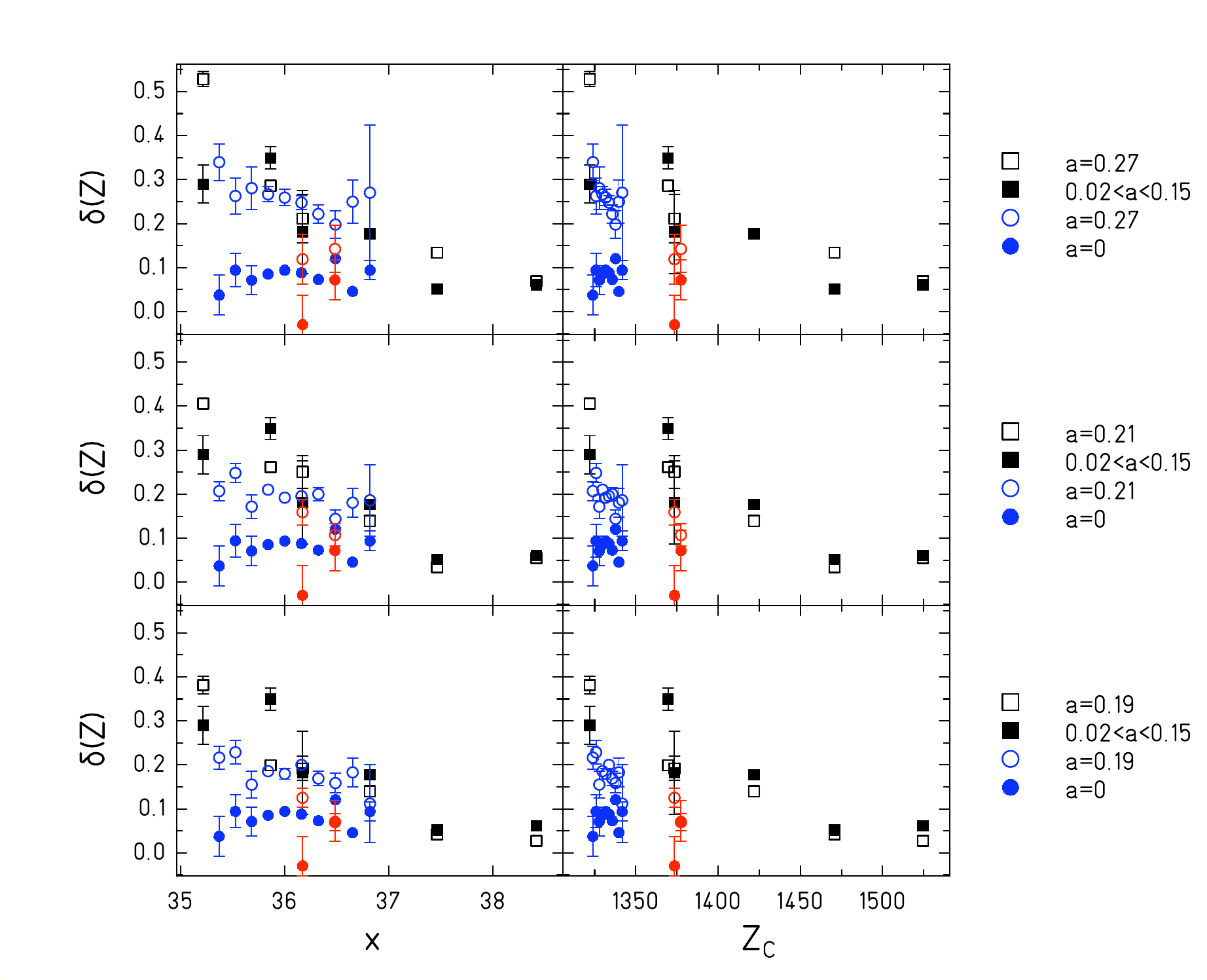}
  \caption{Upper panels : Local even-odd effect as a function of the fissility x (left) and Coulomb parameter $Z_c$ (right), for an asymmetry of 0.27 (open symbols), and compared to the local even-odd effect measured at symmetry (full symbols). Data measured in ielectromagnetic-induced fission on neutron-defficient Th and U isotopes are shown as blue and red circles, respectively. Data on fissile isotopes measured in thermal-neutron-induced fission as black squares. Middle and bottom panels: same as above with asymmetry of 0.21 and 0.19, respectively.}
 \label{symasym}
\end{figure}

The local even-odd effect of the ensemble of the actinides investigated is displayed in figure \ref{symasym}, as a function of the fissility and the Coulomb parameter of the fissioning nucleus, for different values of the asymmetry (open symbols) and compared to the values close to  or at  symmetry (full symbols), depending on the availability of experimental data.  
For symmetric splits measured in electromagnetic-induced fission, the local even-odd staggering shows  small but definitively non-zero amplitude, remarkably independent of the fissioning nucleus. 
This experimental observation extends the previous one put forward in figure \ref{delta_sym} to a wider range of nuclei. The data for fissile isotopes do not provide direct information to this issue, as the symmetry cannot be reached due to technical limitations already mentioned above. The full squares in figure \ref{symasym} correspond to the most symmetric available yields, and due to the properties of the fragment distributions, discussed in section 3, these most symmetric yields are also related to the  fissioning system (decreasing from $a = 0.15$ to $0.02$ when considering fissioning systems from $^{230}$Th to $^{250}$Cf ). This relationship is observed in the steadily decreasing amplitude of the even-odd effect at the most symmetric yield as a function of fissility or Coulomb parameter.

For large asymmetry (upper panels of figure \ref{symasym}), although fission is induced with moderately higher excitation energy in the inverse kinematics technique,  electromagnetic- and thermal-neutron-induced fission data show the same trend. The amplitude of the even-odd staggering at a given asymmetry  is decreasing with fissility or Coulomb parameter. 
When the asymmetry is lowered (shown in the middle and  lower panels of Fig. \ref{symasym}), the slope of the local even-odd effect reduces and gradually approaches the value of the even-odd staggering at symmetry. 
Left and right panels of figure \ref{symasym} show the evolution of the even-odd effect with two different ordering parameters, the fissility and the Coulomb parameter of the fissioning nucleus, respectively. Conventionally \cite{rejmund, naik, bocquet2}, the Coulomb parameter is considered as the most appropriate ordering parameter, since it is associated to the available energy gain on the path from saddle to scission \cite{bouzid}. However, this has always been done within the limited range of accessible nuclei in direct kinematics, and ignoring the correlation between the Coulomb parameter and the average asymmetry of the fragment distribution. Figure \ref{symasym} is displaying data for fixed values of the asymmetry distribution, thus eliminating the role of the asymmetry. Right panels show that  the data of the long isotopic chains of thorium obtained in electromagnetic-induced fission shrink to a narrow range when plotted as a function of the Coulomb parameter. The systematic increase of the even-odd effect with increasing neutron number (established in figure \ref{combi}) results in a steep slope which is proper to each fissioning element; thorium and uranium data follow parallel slopes, which would correspond to a zero value for the even-odd staggering at $Z_c \sim 1350$   and $Z_c \sim 1400$, respectively. However, considerable even-odd effects have been observed for heavier systems
(see figure 1).
Thus, there are severe problems in considering the Coulomb parameter as the ordering parameter for the even-odd effect.This could not be observed earlier, as the number of investigated nuclei was too limited (in general only one isotope per fissioning element, as shown by the black squares), and restricted to nuclei close to the stability line. The other possible ordering parameter, the fissility, is used in the left panels of figure \ref{symasym}. This parameter allows for a coherent description of the even-odd effect throughout the whole ensemble of collected data, originating from electromagnetic- or thermal-neutron-induced fission. The systematic decrease of the even-odd effect at fixed asymmetry suggests that the even-odd effect would disappear at $x  \sim 39$, which is compatible with the available data on even-odd effect in fission, since all systems investigated have smaller values of fissility.


\section{Influence of the excitation energy}

\begin{figure} 
 \includegraphics[height=.3\textheight]{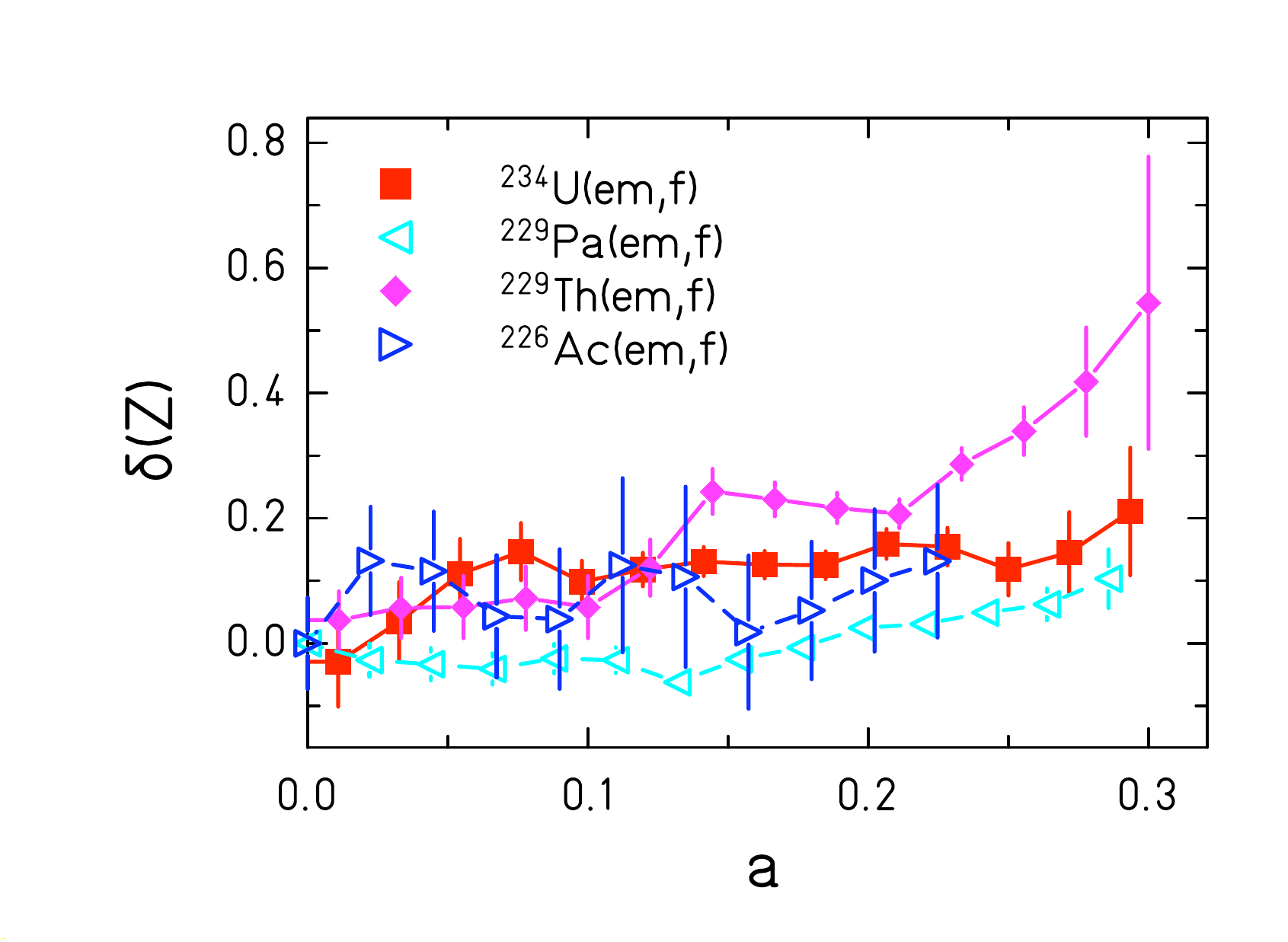}
  \caption{Local even-odd staggering in neighbouring even-Z and odd-Z actinides measured in electromagnetic-induced fission \cite{schmidt}}
 \label{combi-odd}
\end{figure}

Figure \ref{delta_sym} shows that the higher excitation energy influences the even-odd staggering in element yields of the even-Z $^{234}$U and $^{229}$Th systems with a constant reduction factor over the complete fragment distribution. The number of initially broken pairs does have an impact on the amplitude of the even-odd effect. There exists no experimental data on the evolution of the even-odd effect in element distribution of odd-Z fissioning systems with the excitation energy. However, it is possible to compare neighbouring odd-Z and even-Z systems investigated in electromagnetic-induced fission, where the compound nucleus is created with few MeV excitation energy above the fission barrier \cite{schmidt}, which allows for the breaking of several nucleon pairs. Interestingly enough, figure \ref{combi-odd} shows that, in contrast to the low-energy fission case (figure \ref{combi}), at several MeV of excitation energy, the  values for the even-odd staggering in odd-Z fissioning systems are systematically lower than their neighbouring even-Z systems. In low-energy fission, the pairing correlations in even-Z and odd-Z fissioning systems are very different; one is totally paired (the excitation energy at saddle is below the pairing gap) while the second has an ensemble of totally paired protons plus one unpaired proton; this additional proton is expected to weaken strongly the pairing correlations of the nucleus by blocking effects, while figure \ref{combi} shows that the pairing effects in close-by even- and odd-Z fissioning systems have similar amplitude. 
At higher excitation energy,  the proton-proton correlations in both systems are expected to be more similar, as the possible broken pairs in the even- and odd-Z systems equally weaken the pairing correlations of the remaining pairs.  A smaller, but also closer amplitude of the even-odd effect in both types of fissioning systems would have thus been expected at higher excitation energy. This is in contradiction with the data in figure \ref{combi-odd}, which exhibit a systematically lower amplitude of the even-odd effect in odd-Z fissioning nuclei. 
%

\section{Conclusion}

A systematic investigation on even-odd staggering in fission-fragment yields based on thermal-neutron-induced fission of fissile nuclei and electromagnetic-induced fission of neutron-deficient actinides is reported. The present study allows for new properties of the even-odd effect to emerge, which were not accessible previously, due to the limited variety of fissioning nuclei investigated.
The global even-odd effect is shown to vary with properties of the fissioning system, which are not evidently connected to the available energy gain on the deformation path from saddle to scission and the consequent quasi-particle excitations. 
For all considered systems, the amplitude of the local even-odd effect at symmetry is shown to be small compared to the value at large asymmetry, while the latter is varying strongly with the fissioning system. The evolution of the global even-odd effect with the fissioning system is thus shown to be essentially due to the evolution of the fragment-distribution asymmetry, and its influence on the local even-odd effect.
Consequently, the global even-odd effect cannot be considered to derive conclusions on dissipation, as it includes a strong contribution of the asymmetry of the fission-fragment distribution, which is not deriving from energy dissipation. 
The Coulomb parameter of the fissioning system, which is the conventional ordering parameter to investigate the evolution of the even-odd effect with the fissioning nucleus is shown to be questionable. Despite its correlation to the available energy gain in scission path, it is unable to describe the evolution of the even-odd effect over the ensemble of the investigated fissioning systems, which confirms that the even-odd effect is not only connected to the dissipated energy. 

The results of the present analysis refute great part of the traditional theoretical models and ideas, which were developed to explain the even-odd effect in fission-fragment distributions. The new findings should be a guidance for a better understanding of this beautiful signature of the influence of nuclear structure on nuclear dynamics.

 This work has been supported by the R\'egion Basse Normandie with a Chair of Excellence position in GANIL, as well as by the EURATOM programme under the contract number 44816.



\end{document}